%% file: root.tex
\documentclass[letterpaper, 10 pt, conference]{ieeeconf}  

\IEEEoverridecommandlockouts                              

\usepackage{graphics}
\usepackage[pdftex]{graphicx}
\usepackage[tight,footnotesize]{subfigure}
\usepackage{multirow}
\usepackage[cmex10]{amsmath}
\usepackage{amssymb}
\usepackage{amsfonts}
\usepackage{algpseudocode}
\usepackage{algorithm}
\usepackage{booktabs}
\usepackage{url}
\usepackage{gensymb}
\usepackage{color}
\usepackage{hyperref}
\usepackage{comment} 
\usepackage[nocompress]{cite}
\usepackage{xcolor}
\usepackage[table]{xcolor}
\usepackage{makecell}
\definecolor{best}{RGB}{191,232,193}
\definecolor{secondbest}{RGB}{225,241,170}

\title{\LARGE \bf
SIGNN++: Robust Graph Neural Star Identification on Earth
}

\author{Jeongbin Sohn$^{1}$, Jina Lee$^{1}$, Max Yen$^{2}$, Dowan Gwon$^{1}$, Jeahn Han$^{1}$, \\
Floyd Hepburn-Dickins$^{3}$, Brian Coltin$^{4}$, Ryan Soussan$^{4}$, and Pyojin Kim$^{1,\dagger}$%
\thanks{$^{1}$Department of Mechanical and Robotics Engineering, Gwangju Institute of Science and Technology (GIST), Gwangju 61005, South Korea.
        {\tt\small \{sjb010409,\allowbreak jinaleeci,\allowbreak d0wan\_gw0n,\allowbreak jeahnhaan\}@gm.gist.ac.kr, pjinkim@gist.ac.kr}}%
\thanks{$^{2}${\tt\small maxyen@google.com}}%
\thanks{$^{3}${\tt\small frhepburndickins@gmail.com}}%
\thanks{$^{4}$NASA Ames Research Center, CA 94043, USA.
        {\tt\small \{ryan.soussan,brian.coltin\}@nasa.gov}}%
\thanks{$^{\dagger}$Corresponding author.}%
}

\begin{document}

\maketitle
\thispagestyle{empty}
\pagestyle{empty}

\begin{abstract}
We present SIGNN++, a learning-based star identification (SID) framework for celestial navigation in GNSS-denied nighttime environments.
Although researchers have developed learning-based SID methods to improve robustness to observation noise, prior evaluations have focused mainly on synthetic or non-terrestrial environments, leaving their reliability under real terrestrial conditions unclear.
To address these limitations, SIGNN++ extends SIGNN, a graph neural network-based SID method, with three components: a more realistic noise model for training data, Fourier Transform (FT) of angular-distance edge features, and RANSAC-based Perspective Star Alignment for identification and attitude refinement.
The results show that SIGNN++ is more robust to terrestrial observation disturbances and achieves higher star identification accuracy and solved rates than the original SIGNN and publicly available SID baselines, while maintaining comparable attitude accuracy and practical computation time for continuous operation.
The implementation code of SIGNN++ and datasets will be available at {\textcolor{magenta}{\url{https://signnpp.github.io/signnpp/}}}.
\end{abstract}

\input{Introduction}

\input{Related_Work}
\input{Proposed_Method}
\input{Evaluation}
\input{Conclusion}

\bibliographystyle{IEEEtran}
\bibliography{Reference}

\end{document}

%% file: Introduction.tex
\section{Introduction}
\label{sec:Introduction}

Celestial navigation has long been one of the fundamental navigation solutions, guiding ships and aircraft from ancient seafarers to the Age of Exploration on Earth.
Stars still provide an absolute orientation reference that cannot be jammed or spoofed, making them valuable in GNSS-denied military operations and nighttime environments, as Fig.~\ref{fig:celestial_navigation} illustrates.
While spacecraft heavily rely on high-performance star trackers for attitude estimation, terrestrial celestial navigation has received comparatively little attention, and only a few star identification (SID) studies have evaluated attitude estimation on real ground-based datasets.
This work therefore focuses on developing an SID method that operates reliably in terrestrial environments.

\begin{figure}[t]
\label{fig:celestial_navigation}
  \centering
  \includegraphics[width=\columnwidth]{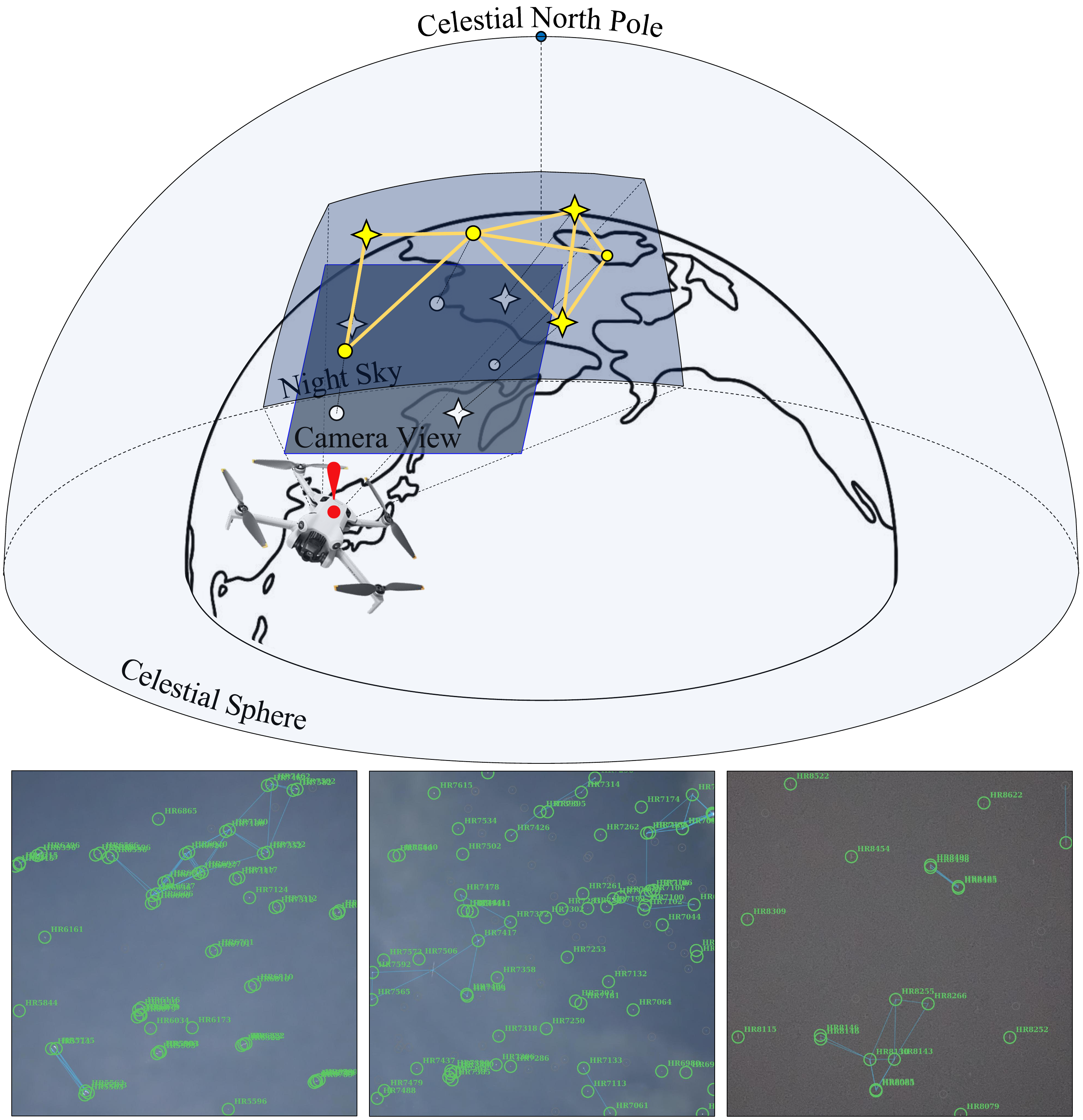}
  \vspace{-6mm}
  \caption{\textbf{Celestial Navigation Overview and SIGNN++ Inference Results.} The system calculates the camera attitude after it correctly identifies the stars in an image. The three images below show, in order, severe motion blur, cloud occlusion, and star streaks. SIGNN++ successfully identifies stars even in terrestrial images with severe noise conditions where conventional SID methods fail.}
  \label{fig:celestial_navigation}
\vspace{-6mm}
\end{figure}

Existing SID algorithms identify stars by constructing constellation-like patterns and comparing them with a precomputed catalog.
However, applying these conventional identification methods in terrestrial environments is difficult because of atmospheric refraction, cloud occlusion, and other sources of noise.
Some other work has developed various deep-learning-based SID algorithms, including SIGNN \cite{hepburn2025signn}, which models stars as a graph using a GNN.
However, SIGNN lacks validation on real-world images and shows limited robustness to combined noise similar to that in real environments.
In addition, its magnitude-based node feature heavily depends on photometric calibration.

To address these issues, we propose SIGNN++ by rebuilding the training dataset to mimic terrestrial observation noise as closely as possible and by applying FT to the edge-distance features, which are less sensitive to photometric variation\cite{spratling2009survey, rijlaarsdam2020survey}.
We jointly optimize the inferred IDs and camera attitude to obtain the final star IDs and camera attitude.
We also capture real-world images while rotating the camera on a 3-axis rotation stage and obtain ground-truth (GT) with a Rocket Lab ST-RT~\cite{rocketlab_strt} star tracker to evaluate accuracy and robustness in terrestrial observations.

Our main contributions are as follows:
\begin{itemize}
\item We propose SIGNN++, which extends SIGNN to real terrestrial skies via noise-aware training, Fourier edge features, and a perspective star alignment (PSA).

\item We validate SIGNN++ on real ground-based night-sky images using a co-mounted Rocket Lab ST-RT star tracker as the 3-DoF attitude reference.

\item We publicly release a dataset of timestamped night-sky images paired with star tracker attitude measurements.
\end{itemize}

%% file: Related_Work.tex
\section{Related Work}
\label{sec:Related_Work}

\begin{figure*}[t]
  \centering
  \includegraphics[width=\textwidth]{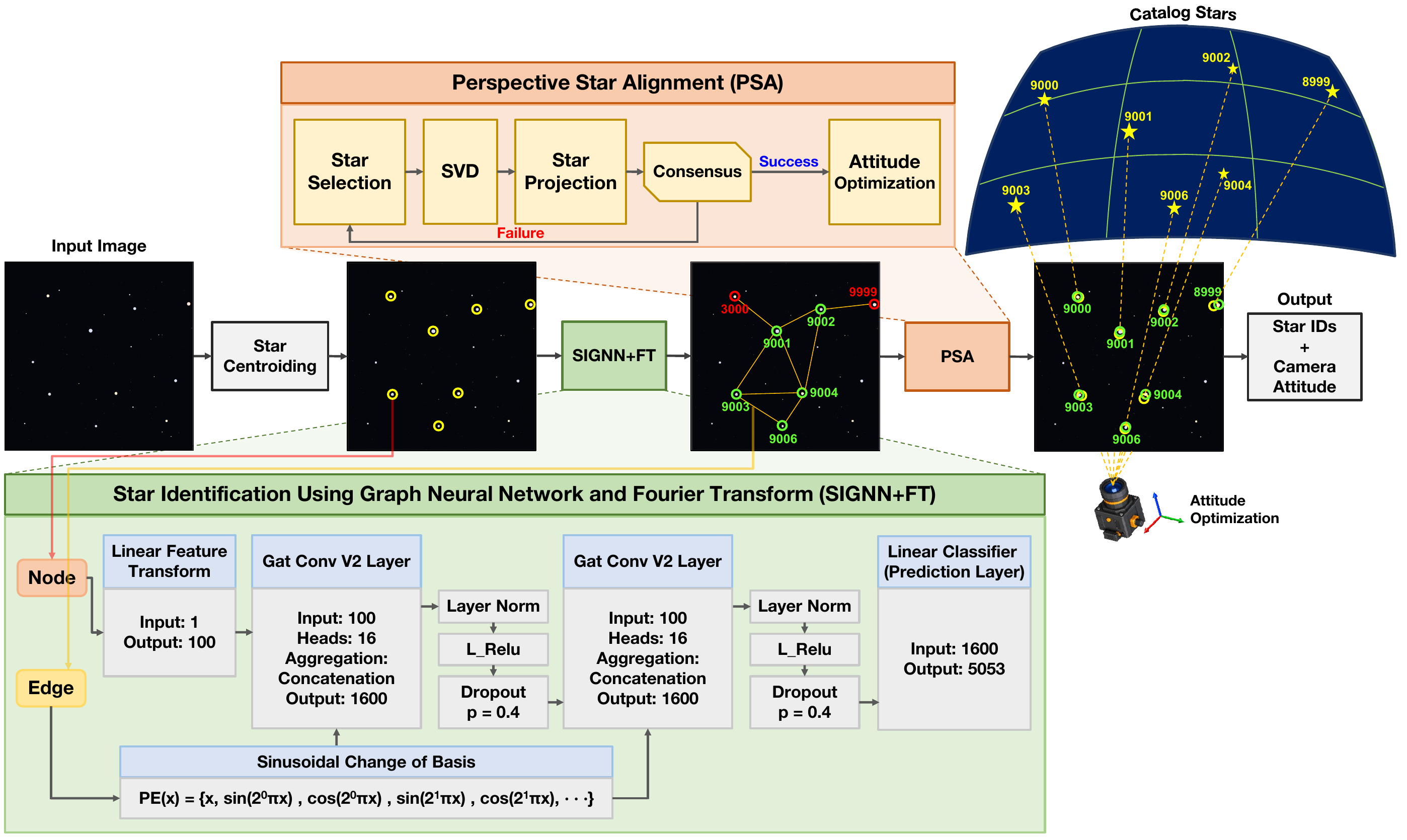}
  \vspace{-6mm}
  \caption{\textbf{SIGNN++ Architecture.} It inputs a night sky image and outputs star IDs and the camera's attitude. We use the center-of-gravity (CoG) method for star centroiding.
  Since centroiding is not the focus of this work, we use the CoG method as a simple and widely adopted baseline and provide identical centroid inputs to all SID methods for fair comparison. \ref{subsec:fourier} and \ref{subsec:psa} describe the FT module and Perspective Star Alignment, respectively.}
  \label{fig:pipeline_architecture}
\vspace{-6mm}
\end{figure*}

\textbf{Deterministic SID methods} identify stars by matching geometric relations between observed and catalog stars. Representative approaches include Grid, Pyramid, Log-polar, Tetra, and ROSIA \cite{padgett2002grid,mortari2004pyramid,wei2009star,brown2017tetra,Chng_2023}. 
Although these methods can achieve high accuracy under clean observations, their reliance on geometric consistency makes them sensitive to positional errors, false detections, and other disturbances that frequently occur in terrestrial night-sky images.
Among them, we use Tetra and ROSIA as deterministic baselines because their publicly available implementations enable reproducible comparison.

\textbf{Learning-based SID methods} are developed to improve robustness to observation noise by replacing or augmenting hand-crafted geometric descriptors with learned representations \cite{wang2021efficient,Luo:26,8764526}. However, prior studies evaluate these methods mainly in synthetic or non-terrestrial environments. SIGNN \cite{hepburn2025signn} models stars as a graph and uses noise-augmented training to improve robustness, showing strong performance in synthetic noise ablations. Nevertheless, SIGNN does not report evaluation on real terrestrial star images, and its performance degrades under multiple simultaneous noise types. SIGNN++ directly addresses these limitations by using a more realistic terrestrial noise model, Fourier-transformed edge features, and Perspective Star Alignment for joint identification and attitude refinement.

\textbf{Terrestrial stellar navigation} has been investigated using deterministic matching, smartphone-based star tracking, fixed-wing UAVs, and event cameras \cite{s20041106,drones8110652,reed2025ebsekfaccuratehighfrequency}. However, these systems primarily rely on deterministic initialization, external attitude systems, or subsequent tracking rather than end-to-end evaluation of a learning-based SID method on real terrestrial observations. In contrast, this work evaluates a learning-based SID pipeline from star identification to attitude estimation on real terrestrial night-sky images.

%% file: Proposed_Method.tex
\section{Proposed Method}
\label{sec:Proposed_Method}
SIGNN++ extends the original SIGNN for terrestrial observations by introducing a terrestrial noise-aware training dataset, Fourier-transformed edge features, and a RANSAC-based Perspective Star Alignment (PSA) module for attitude estimation and ID refinement. Fig.~\ref{fig:pipeline_architecture} illustrates the overall workflow. Given a night-sky image, SIGNN++ extracts star centroids, performs SIGNN-based star identification, and subsequently refines the predicted IDs and camera attitude through PSA. The detailed procedures of Fourier features and PSA are described in Sections~\ref{subsec:fourier} and \ref{subsec:psa}, respectively.

\subsection{SIGNN}
\label{subsec:workflow}
SIGNN represents detected stars as a graph, where stars are defined as nodes and the geometric relationships between stars are represented as edges. This graph representation is suitable for star identification because it directly captures the geometric topology of star patterns while ignoring irrelevant background regions. In addition, the graph structure naturally provides rotation invariance, enabling consistent identification regardless of image orientation.

SIGNN processes the star graph using a 2-layer Graph Attention Network (GAT). The GAT aggregates information from a star's direct neighbors as well as its 2-hop neighbors through message passing, allowing the network to learn local geometric relationships between stars. The node feature is the relative magnitude between neighboring stars, and the edge feature is the angular distance between stars, which is calculated by converting image pixel coordinates into celestial coordinates using a pinhole camera model. After the GAT layers, the aggregated node representations are passed to a linear classifier to predict the final star IDs. Further details are provided in \cite{hepburn2025signn}.

\vspace{-2mm}
\subsection{Fourier Transform in SIGNN}
\label{subsec:fourier}
The FT module employs Fourier features to map the one-dimensional angular distance into a high-dimensional space, which enables the GNN to perceive fine-grained geometric structures. By transforming scalar distances using a set of sinusoidal functions with varying frequencies, the module mitigates the spectral bias of deep networks \cite{rahaman2019spectralbiasneuralnetworks, tancik2020fourierfeaturesletnetworks}. Consequently, encoding GNN edge distances with high-frequency Fourier features increases the discriminability of geometrically similar edges, providing more distinctive edge representations for star identification. Eq.~(\ref{eq:positional_encoding}) defines the Fourier feature mapping.

The module computes the features as follows:

\vspace{-2mm}
\begin{equation}
\label{eq:positional_encoding}
    \gamma(x) = \left( x,\ \left( \sin(2^k \pi x),\, \cos(2^k \pi x) \right)_{k=0}^{L-1} \right)
\vspace{-2mm}
\end{equation}

where $x$ is the angular distance and $L$ is a hyperparameter. In this study, we set $L = 6$, which provides the best performance.

\subsection{Perspective Star Alignment (PSA)}
\label{subsec:psa}
SIGNN++ passes the predicted star IDs to Perspective Star Alignment (PSA), which estimates the camera attitude and aggregates inliers by comparing with the star catalog.
SIGNN+FT outputs an ID and confidence for each detected star, from which PSA randomly samples two stars from the 24 highest-confidence predictions to estimate an initial rotation \(R\) using singular value decomposition (SVD)~\cite{wahba}.
PSA projects the detected stars onto the celestial sphere using \(R\) and counts stars with a catalog star within \(0.5^\circ\) as candidate inliers.
If the number of inliers exceeds a predefined threshold, PSA samples new star pairs from the inlier set to re-estimate \(R\) and updates the rotation~\cite{FISCHLER1987726}.

After inlier aggregation, iterative closest point (ICP) establishes one-to-one correspondences between the detected centroids and catalog stars.
A nonlinear least-squares solver then refines the attitude by jointly optimizing the field of view (FOV) and rotation $R$.
The refinement minimizes the chordal residual between the rotated image rays and matched catalog directions using a Trust-Region-Reflective least-squares solver with a soft-L1 loss.
Motivated by the iterative attitude refinement used in Tetra~\cite{brown2017tetra}, PSA formulates this process within a RANSAC framework.
This enables SIGNN++ to obtain accurate camera attitude while achieving identification precision comparable to Tetra.

\subsection{Image-Based Star Magnitude Estimation}
The original SIGNN assumes that the stellar magnitudes are known, which is a strong assumption for practical SID with real images. To relax this assumption, we compute the apparent magnitude of each star by following the conventional photometric procedure as closely as possible. First, the input image is normalized, and bright pixels are grouped to determine the star centroids using the CoG method. The original image is then converted to a luminance map, where different weights are assigned to the RGB channels according to the sensitivity of the human eye. We define an aperture and an annulus around each centroid, and we compute the flux as the sum of the pixel values within the aperture minus the median pixel value in the annulus multiplied by the aperture area. We then calculate the apparent magnitude using Pogson's equation \cite{h_karttunen_fundamental_2016}:

\vspace{-2mm}
\begin{equation}
\label{eq:magnitude}
    M = -2.5 \log_{10}(F) + C
\vspace{-2mm}
\end{equation}

where \(M\) denotes the apparent magnitude, \(F\) denotes the measured flux, and the reference flux determines the constant \(C\). In this study, we empirically set \(C\) to 12, although this value may vary depending on the sensor type and atmospheric conditions. SIGNN uses the relative magnitude between neighboring stars as a node feature and is therefore relatively robust to variations in \(C\). In contrast, ROSIA uses absolute magnitude information and is more sensitive to this value \cite{Chng_2023}. Despite estimation errors, the computed magnitude provides useful node information for SIGNN.

\begin{figure}[h] 
  \centering
  \includegraphics[width=\linewidth]{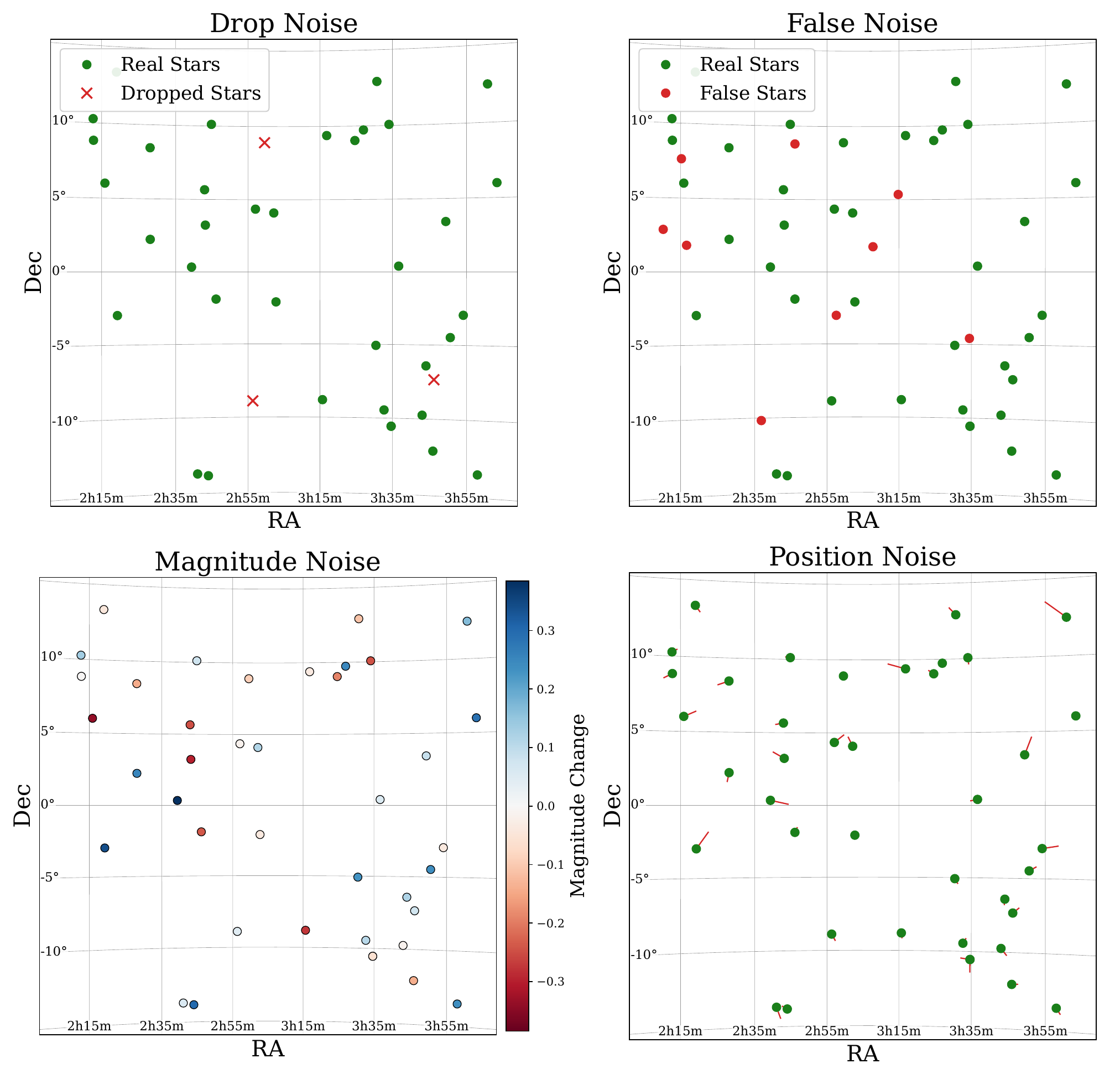}
  \vspace{-8mm}
  \caption{\textbf{Visualization of four types of training noise.} Drop and false-star noise remove and add stars with fixed probabilities, respectively, while magnitude and positional noise apply perturbations sampled from Gaussian distributions.}
  \label{fig:training_noise}
\vspace{-4mm}
\end{figure}

\subsection{Noise Application in the Simulated Dataset}
For training SIGNN++, we apply the same four noise types that previous learning-based SID methods such as SIGNN and RPNet use: false-star, drop, magnitude, and positional noise, while attempting to reproduce how these noise sources occur in terrestrial observations.
Most importantly, real images do not contain only one type of noise at a time.
Therefore, we independently apply each noise type with a probability of 0.7. At the same time, 20\% of the frames are kept noise-free to provide a reference for learning the underlying geometric structure, where we initially assume that all stars brighter than magnitude 6 are detected. Fig. \ref{fig:training_noise} visualizes the four types of training noise.

We design drop noise so that dimmer stars have a higher probability of disappearing, in order to mimic the phenomenon that faint stars become difficult to observe under severe light pollution in terrestrial night-sky images. We model the drop probability using a logistic curve. For false stars, the magnitude is not sampled uniformly from 0 to 6. Instead, we generate dimmer false stars more frequently, following an approximate stellar-count slope in which the number of stars increases by a factor of 2.5 for each increase of one magnitude. In real observations, faint false detections caused by clouds, birds, and similar objects are more common than very bright false objects. We therefore assume that this distribution better reflects real observations. The number of false stars is determined by a ratio sampled from a uniform distribution over \([0,\text{noise level}]\), and the false stars are inserted at random image locations.

We apply magnitude and positional noise in the same manner as SIGNN but increase their maximum noise levels tenfold, to 1 and 0.01, respectively.
For both noise types, we sample a standard deviation from a uniform distribution over \([0,\text{noise level}]\) and then apply Gaussian noise with the sampled standard deviation to each star.
As a result, 14,000 training images are generated per epoch, with each image corresponding to a random view on the celestial sphere using a circular FOV cap with a diameter of \(25^\circ\).
In this way, the training dataset for graph learning is designed to resemble the real terrestrial images used as the evaluation dataset as closely as possible.

%% file: Evaluation.tex
\section{Evaluation}
\label{sec:Evaluation}

The evaluation consists of three parts: noise-ablation experiments on synthetic data, terrestrial simulation experiments using Stellarium, and real-world experiments using random-view and continuous night-sky images with star-tracker-based ground truth.

For ROSIA, we adjust the angular inlier radius to \(2\times10^{-3}\) rad and the magnitude gate to 5.0 because the original parameters were optimized for simulated scenes and showed limited performance on real-image experiments.

\subsection{Dataset Generation}
We adopt a three-tiered evaluation protocol using three distinct datasets. We directly feed all images into the SID pipeline without additional calibration.

\noindent\textbf{Catalog Dataset}\quad
We use the catalog dataset to train and validate SIGNN++. Stars brighter than magnitude 6 are selected from the Bright Star Catalog 5 (BSC5) \cite{hoffleit1991bsc}, and each sample is constructed using stars within a circular FOV with a diameter of $25^\circ$. The applied noise and camera viewing direction are changed at every epoch, making the effective size of the training dataset virtually unlimited. For the ablation study, the diameter is $25^\circ$, the image resolution is $2048\times2048$ pixels, and we use only samples containing at least ten stars. We apply noise following the SIGNN~\cite{hepburn2025signn} procedure, except that we resample the magnitudes of false stars from the observed magnitude distribution of the corresponding frame rather than generating them randomly. A total of 100 samples are generated for each noise level, and only stars within a $10^\circ$ radius from the FOV center are evaluated. Pooled recall is used as the evaluation metric.

\begin{figure}[h] 
  \centering
  \includegraphics[width=\linewidth]{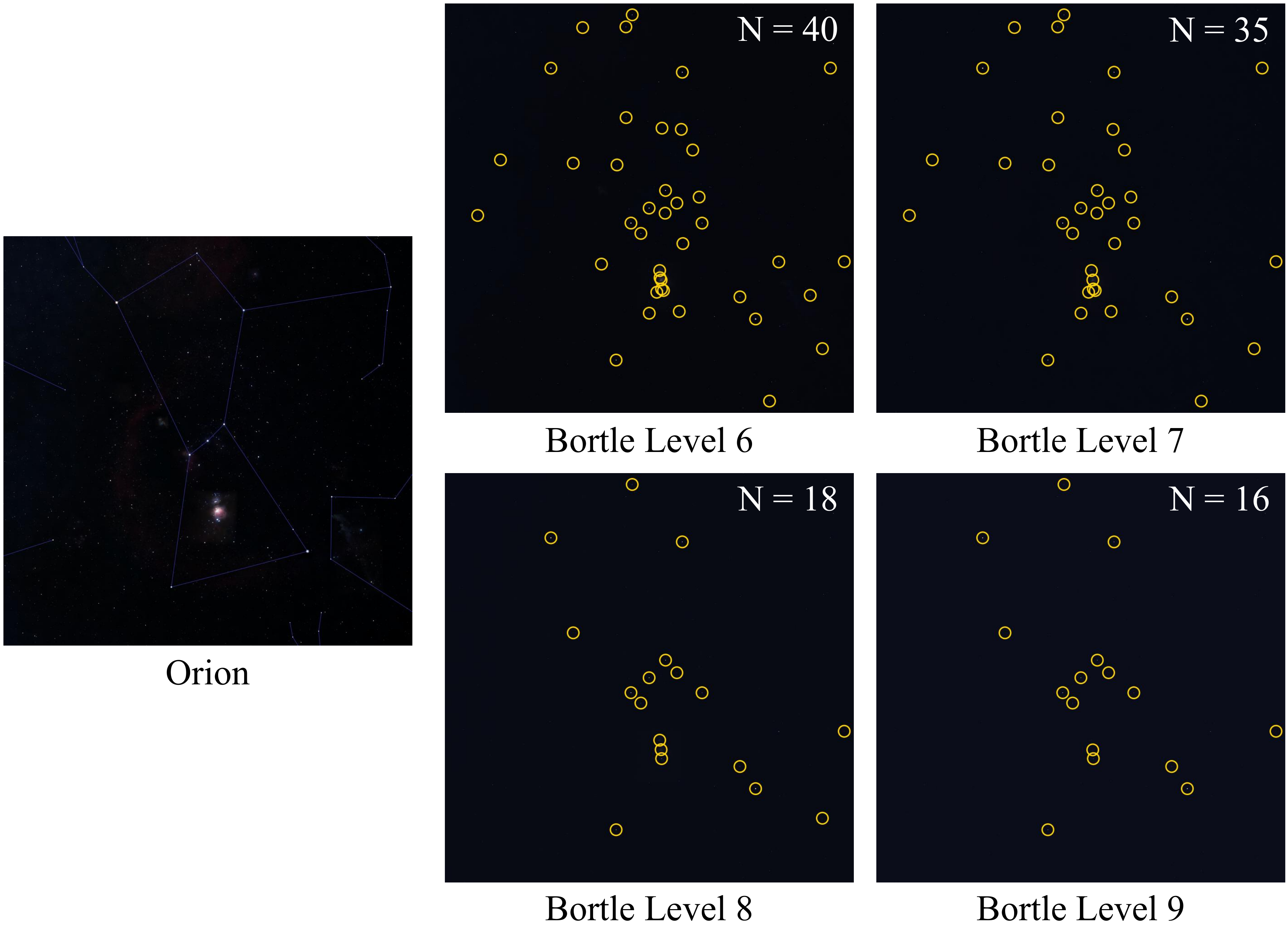}
  \vspace{-8mm}
  \caption{\textbf{Visualization of four levels of light pollution rendered by Stellarium.} The left image shows the Orion constellation observed without atmospheric visualization, while the right images show star centroiding results under atmospheric effects with different Bortle levels for the same scene. As the Bortle level increases, fewer stars are detected.}
  \label{fig:bortle_level}
\vspace{-6mm}
\end{figure}

\begin{figure}[h] 
  \centering
  \includegraphics[width=0.9\linewidth]{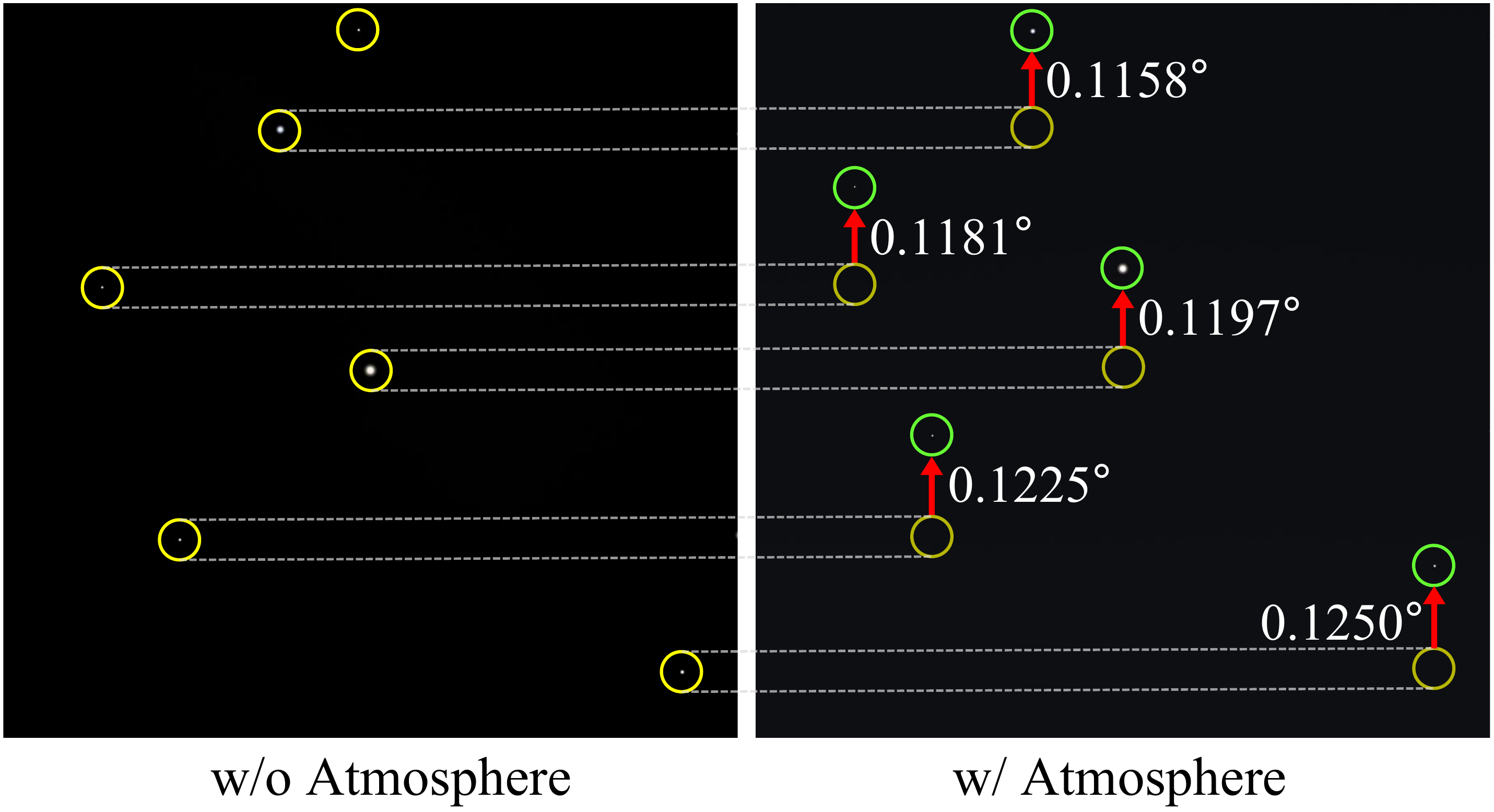}
  \vspace{-2mm}
  \caption{\textbf{Visualization of atmospheric refraction rendered by Stellarium.} Near the horizon, atmospheric refraction introduces significant positional errors in star centroids.}
  \label{fig:refraction}
\vspace{-2mm}
\end{figure}

\noindent\textbf{Stellarium Dataset}\quad
We construct the Stellarium dataset because Stellarium enables observations over the entire celestial sphere while providing control over atmospheric conditions.
We model light pollution using four Bortle levels from six to nine, corresponding to bright suburban sky, suburban/urban transition, city sky, and inner-city sky, respectively (see Fig.~\ref{fig:bortle_level}).
It is also divided according to horizontality: the sky condition uses elevation angles from $50^\circ$ to $90^\circ$, while the horizontal condition uses elevation angles from $10^\circ$ to $30^\circ$.
The atmospheric refraction effect according to the observation elevation is illustrated in Fig.~\ref{fig:refraction}.
The viewing direction is randomly selected, the FOV is $20^\circ$, we use stars brighter than magnitude 6, and each image contains at least ten visible stars.
In total, 5,000 images are captured, with 625 images generated for each combination of Bortle level and horizontality.

\noindent\textbf{Real-World Dataset}\quad
The real dataset is divided into two types: random-view images for comparing identification performance and continuous sequences for evaluating real-time capability and suitability for terrestrial environments. For the random-view dataset, we capture \(2000 \times 2000\)-pixel images with a \(20^\circ\) FOV and a nominal exposure time of 1.6~s using the night mode of an iPhone 15 Pro Max on July 15, July 20, and July 28, with 19, 14, and 34 images collected on each date, respectively. A considerable number of these images contain clouds and motion blur. The iPhone 15 Pro Max is selected as a representative commercial off-the-shelf (COTS) camera to evaluate camera-based SID performance under practical terrestrial conditions, rather than relying on expensive dedicated astronomical cameras. To evaluate identification and attitude, we obtain reference star IDs and attitudes using Astrometry.net, which ROSIA also uses to obtain reference star IDs and attitudes \cite{Chng_2023}.

\begin{figure}[h] 
  \centering  \includegraphics[width=0.9\linewidth]{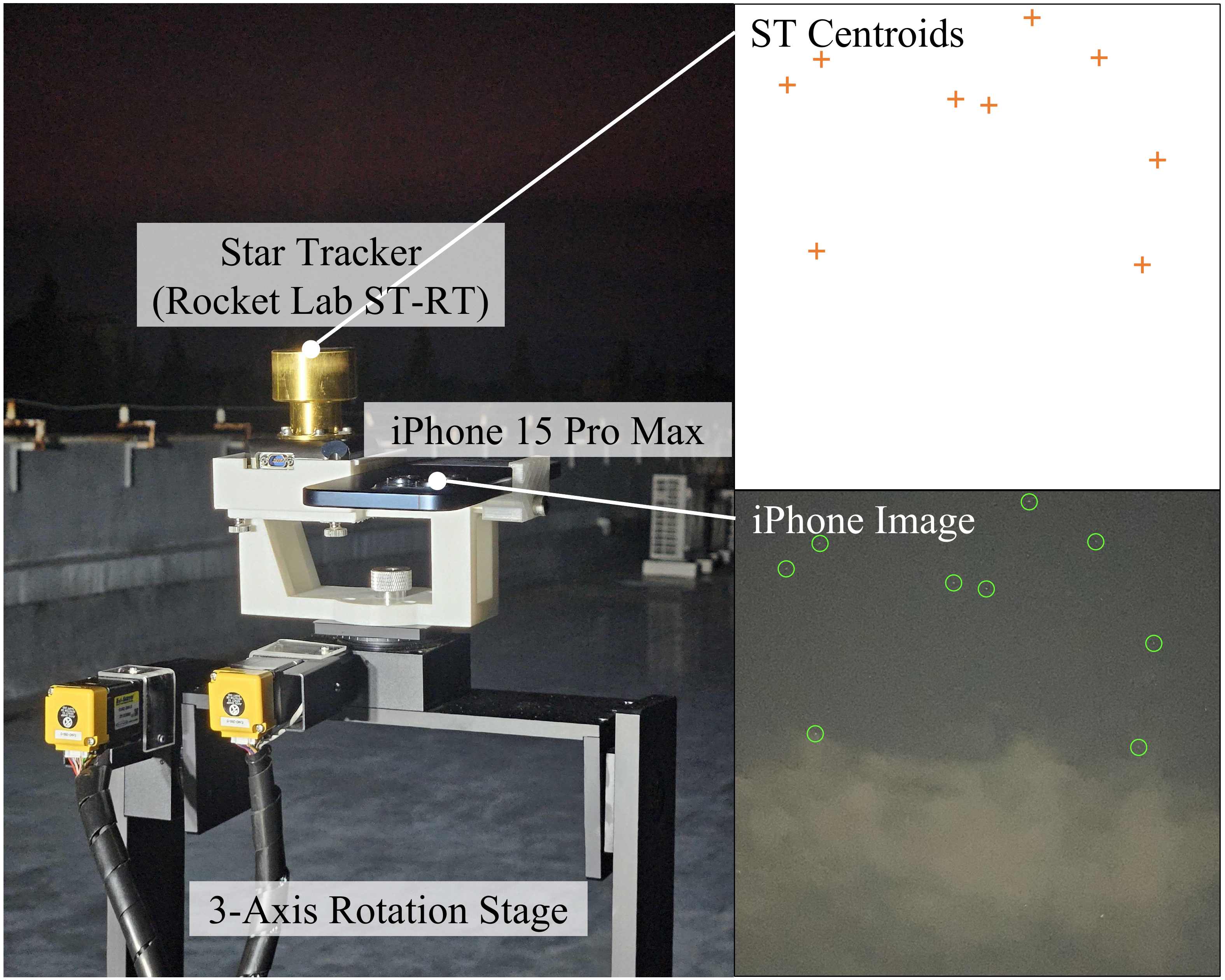}
  \vspace{-2mm}
  \caption{\textbf{Experimental setup for acquiring sequential real sky images.} The setup consists of a rotation stage, an iPhone, and a star tracker (ST). The iPhone image on the right is an actual image captured with a 500~ms exposure, whereas the ST centroid image is an illustration showing the expected centroid locations of stars rather than an actual ST image.}
  \label{fig:experimental_setup}
\vspace{-4mm}
\end{figure}

For the continuous-sequence dataset, an ST-RT star tracker (ST) from Rocket Lab \cite{rocketlab_strt} and an iPhone 15 Pro Max with a \(20^\circ\) FOV are mounted on a rotation stage and rotated at a constant angular velocity. The ST-RT is used as the reference to obtain attitude measurements. Each sequence consists of eight to ten images with a resolution of \(2000\times2000\) pixels. Fig.~\ref{fig:experimental_setup} shows the experimental setup. We capture a total of 15 sequences on September 10 at 0.5~fps with an exposure time of 500~ms. The sequences are collected under various imaging elevations, rotation directions, and angular velocities. Table~\ref{tab:sequence_performance} shows the setup for each sequence. Since the latency of the ST-RT is not always constant, the variable ST-RT latency prevents perfect synchronization between the ST-RT and the camera.

Because of a synchronization mismatch between the ST and the iPhone camera, we calibrate the attitude error using an offset equal to the median attitude over the entire sequence and do not calculate the attitude error for sequences in which the method identifies only one frame. 

\subsection{Evaluation Metrics}
We evaluate SID performance using four main metrics. First, identification accuracy is an important metric, but its definition varies across the literature \cite{hepburn2025signn, 8764526, padgett2002grid, Chng_2023}.
Since precision is critical for accurate attitude estimation, while recall is also important for evaluating pure identification performance, this paper reports precision, recall, and F1 score for a fair evaluation. We define the solved rate as the proportion of frames for which a method identifies at least three stars or, when GT star IDs are unavailable, as the proportion of frames with an attitude error below \(1^\circ\). Three stars are sufficient to determine a unique attitude solution. Next, computation time is defined as the time required for star identification and attitude estimation, excluding centroiding. We measure and report the computation time of all methods using an AMD Ryzen Threadripper 7960X workstation. Finally, the attitude estimation error is computed using the absolute rotation error introduced in \cite{kim2018indoor}, which measures the difference between rotation matrices. We obtain the ground-truth attitude for Stellarium images from Stellarium's internal information and the ground-truth attitude for real images from Astrometry.net or an ST.

\subsection{Catalog Dataset}
\begin{figure}[h] 
  \centering
  \includegraphics[width=\linewidth]{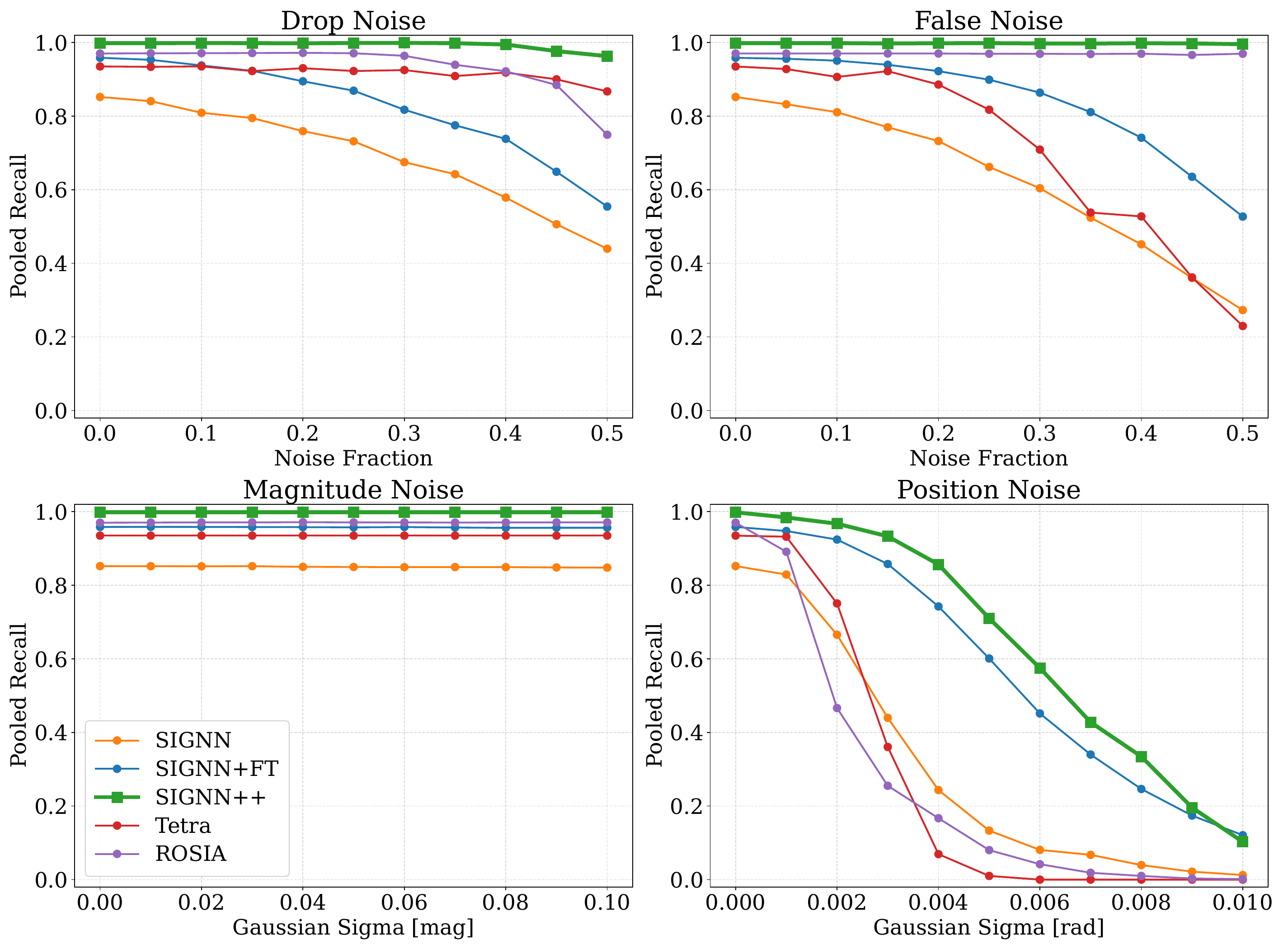}
  \vspace{-8mm}
  \caption{\textbf{Ablation on the four noises on the catalog dataset.} The figure compares SIGNN++ with SIGNN \cite{hepburn2025signn}, Tetra \cite{brown2017tetra}, and ROSIA \cite{Chng_2023}. Each ablation consists of 11 noise levels, including the noise-free case. The maximum noise levels are 50\% for drop noise, 50\% for false-star noise, 0.1 for magnitude noise, and 0.01~rad for positional noise.}
  \label{fig:noise_ablation}
\vspace{-2mm}
\end{figure}

Noise ablation is conducted to evaluate the robustness of each algorithm against individual noise types. As shown in Fig.~\ref{fig:noise_ablation}, SIGNN++ achieves the best overall performance. The overall robustness ranking is SIGNN++, Tetra, SIGNN+FT, ROSIA, and SIGNN. The maximum positional noise reaches 0.01~rad, which corresponds to approximately \(0.5^\circ\).

As expected, Tetra is particularly vulnerable to positional noise because it relies on angular distance information for star pattern matching. In addition, Tetra performance degrades as the drop and false-star noise levels increase because it fails when bright stars are removed or bright false stars are introduced. Since Tetra does not use magnitude information, it is not affected by magnitude noise, which is also observed in the ablation results.

FT improves the robustness of SIGNN across overall noise types, with a particularly noticeable improvement under positional noise. This is because applying FT to the edge features enables the network to learn higher-frequency components of the geometric representation, resulting in more robust identification against positional perturbations.

Magnitude-noise results remain nearly unchanged for SIGNN and ROSIA because the catalog dataset provides accurate stellar magnitudes.

\begin{table}[h]
\vspace{-2mm}
\centering
\caption{Performance comparison across Bortle levels.}
\vspace{-2mm}
\label{tab:bortle_performance}
\scriptsize
\setlength{\tabcolsep}{2pt}

\begin{tabular}{@{}lccccc@{}}
\toprule
\textbf{Method} & \textbf{Metric} &
\textbf{Bortle 6} & \textbf{Bortle 7} & \textbf{Bortle 8} & \textbf{Bortle 9} \\
\midrule

\multirow{3}{*}{\textbf{SIGNN}}
& Solved Rate (\%)
& \cellcolor{secondbest}84.16
& \cellcolor{secondbest}85.28
& 57.78
& \textcolor{red}{19.72} \\
& F1 (\%)
& 50.37 & 51.90 & 38.85 & 32.79 \\
& Comp. Time (s)
& \cellcolor{best}0.01
& \cellcolor{best}0.01
& \cellcolor{best}0.01
& \cellcolor{best}0.01 \\
\midrule

\multirow{4}{*}{\textbf{SIGNN++}}
& Solved Rate (\%)
& \cellcolor{best}96.88
& \cellcolor{best}95.68
& \cellcolor{best}86.36
& \cellcolor{best}64.59 \\
& F1 (\%)
& \cellcolor{best}98.94
& \cellcolor{best}98.98
& \cellcolor{best}99.06
& \cellcolor{best}98.69 \\
& Comp. Time (s)
& \cellcolor{secondbest}0.29
& \cellcolor{secondbest}0.29
& \cellcolor{secondbest}0.14
& \cellcolor{secondbest}0.05 \\
& Attitude Error (deg)
& \cellcolor{secondbest}0.04
& \cellcolor{secondbest}0.03
& \cellcolor{secondbest}0.03
& \cellcolor{best}0.03 \\
\midrule

\multirow{4}{*}{\textbf{ROSIA}}
& Solved Rate (\%)
& 82.00 & 82.24 & 53.45 & \textcolor{red}{16.35} \\
& F1 (\%)
& 95.02 & 94.64 & 91.31 & 85.40 \\
& Comp. Time (s)
& 0.84 & 0.60 & 0.70 & 0.63 \\
& Attitude Error (deg)
& 0.29 & 0.30 & 0.38 & 0.67 \\
\midrule

\multirow{4}{*}{\textbf{Tetra}}
& Solved Rate (\%)
& 79.52
& 79.76
& \cellcolor{secondbest}60.35
& \cellcolor{secondbest}\textcolor{red}{21.45} \\
& F1 (\%)
& \cellcolor{secondbest}98.02
& \cellcolor{secondbest}98.03
& \cellcolor{secondbest}97.63
& \cellcolor{secondbest}97.01 \\
& Comp. Time (s)
& 0.39 & 0.32 & 0.29 & 0.27 \\
& Attitude Error (deg)
& \cellcolor{best}0.03
& \cellcolor{best}0.03
& \cellcolor{best}0.03
& \cellcolor{secondbest}0.03 \\

\bottomrule
\end{tabular}
\vspace{-4mm}
\end{table}

\subsection{Stellarium Dataset}
The Stellarium evaluation is reported using two tables. Table~\ref{tab:bortle_performance} presents the frame solved rate, F1 score, computation time, and attitude error of each algorithm across the Bortle levels. For each metric, the best-performing value is highlighted in green, the second-best value is highlighted in light green, and particularly poor values are written in red. We use the same highlighting scheme for the following tables. For SIGNN++, ROSIA, and Tetra, which perform attitude estimation, the attitude error is also reported.

SIGNN++ achieves the highest solved rate and F1 score at all Bortle levels, showing it has a high probability of correctly identifying at least three stars from arbitrary viewing directions even under severe light pollution.
In addition, while the performance of SIGNN, Tetra, and ROSIA drops sharply at Bortle level 9, SIGNN++ shows only a limited decrease. This robustness is likely attributed to the training dataset design of SIGNN++, where faint stars are more likely to be removed during drop-noise augmentation, effectively mimicking the star loss caused by severe light pollution.

SIGNN achieves the shortest computation time, while SIGNN++ requires additional computation for attitude optimization.
For attitude error, Tetra achieves the lowest error at Bortle levels 6, 7, and 8, while SIGNN++ achieves the lowest error at Bortle level 9.
However, the differences between the two algorithms are less than 0.001\degree, which is insignificant.

ROSIA does not achieve the best value for any metric, which may be partly attributed to errors in apparent-magnitude estimation.
Even its advantage in computation time becomes less significant in this evaluation.
This is because ROSIA explores more candidates when the magnitude information is unreliable, resulting in increased computational cost.
Overall, SIGNN++ shows greater robustness than the other methods as the level of light pollution increases.

\begin{table}[h]
\vspace{-2mm}
\centering
\caption{Performance comparison according to horizontality.}
\vspace{-2mm}
\label{tab:horizontality_performance}
\setlength{\tabcolsep}{4pt}

\begin{tabular}{lccc}
\toprule
\textbf{Method} & \textbf{Metric} & \textbf{Sky} & \textbf{Horizon} \\
\midrule

\multirow{3}{*}{\textbf{SIGNN}}
& Solved Rate (\%)
& 76.00
& \cellcolor{secondbest}47.87 \\
& F1 (\%)
& $51.140 \pm 18.444$
& $39.921 \pm 15.588$ \\
& Comp. Time (s)
& \cellcolor{best}$0.012 \pm 0.003$
& \cellcolor{best}$0.011 \pm 0.003$ \\
\midrule

\multirow{4}{*}{\textbf{SIGNN++}}
& Solved Rate (\%)
& \cellcolor{best}94.87
& \cellcolor{best}77.06 \\
& F1 (\%)
& \cellcolor{best}$99.065 \pm 2.072$
& \cellcolor{best}$98.773 \pm 2.817$ \\
& Comp. Time (s)
& \cellcolor{secondbest}$0.271 \pm 0.263$
& \cellcolor{secondbest}$0.125 \pm 0.201$ \\
& Att. Error (deg)
& \cellcolor{secondbest}$0.026 \pm 0.032$
& \cellcolor{secondbest}$0.035 \pm 0.034$ \\
\midrule

\multirow{4}{*}{\textbf{ROSIA}}
& Solved Rate (\%)
& 75.12 & 42.28 \\
& F1 (\%)
& $95.085 \pm 9.835$
& $90.323 \pm 14.513$ \\
& Comp. Time (s)
& $0.640 \pm 0.697$
& $0.823 \pm 0.905$ \\
& Att. Error (deg)
& $0.292 \pm 0.530$
& $0.427 \pm 0.613$ \\
\midrule

\multirow{4}{*}{\textbf{Tetra}}
& Solved Rate (\%)
& \cellcolor{secondbest}77.40
& 43.45 \\
& F1 (\%)
& \cellcolor{secondbest}$97.925 \pm 2.241$
& \cellcolor{secondbest}$97.684 \pm 2.585$ \\
& Comp. Time (s)
& $0.316 \pm 0.577$
& $0.354 \pm 0.651$ \\
& Att. Error (deg)
& \cellcolor{best}$0.025 \pm 0.014$
& \cellcolor{best}$0.031 \pm 0.033$ \\

\bottomrule
\end{tabular}
\vspace{-2mm}
\end{table}

Table~\ref{tab:horizontality_performance} presents the performance metrics of each algorithm according to horizontality. The comparison SID algorithms show relatively low solved rates near the horizon, whereas SIGNN++ achieves a solved rate of approximately 77\%. This discrepancy is likely caused by the positional error induced by atmospheric refraction, as shown in Fig.~\ref{fig:refraction}. The positional error reaches up to approximately 0.002~rad in the illustrated example; however, the noise ablation results show that SIGNN++ can tolerate this level of positional noise. Therefore, SIGNN++ is expected to remain robust even in horizon observations where severe positional distortion occurs. This indicates that SIGNN++ is also robust in environments with severe atmospheric distortion.

\begin{table}[h]
\vspace{-2mm}
\centering
\caption{Performance comparison on the real random-view dataset.}
\vspace{-2mm}
\label{tab:real_random_view}
\footnotesize
\setlength{\tabcolsep}{2.5pt}

\begin{tabular}{@{}lccccc@{}}
\toprule
\makecell[c]{\textbf{Method}} &
\makecell[c]{\textbf{Solved}\\\textbf{Rate (\%)}} &
\makecell[c]{\textbf{F1 (\%)}} &
\makecell[c]{\textbf{Time}\\\textbf{Solved (s)}} &
\makecell[c]{\textbf{Time}\\\textbf{Failed (s)}} &
\makecell[c]{\textbf{Attitude}\\\textbf{Error (deg)}} \\
\midrule

\textbf{SIGNN}
& 59.7
& 32.63
& \cellcolor{best}0.012
& \cellcolor{secondbest}0.012
& -- \\

\textbf{SIGNN++}
& \cellcolor{best}98.5
& \cellcolor{best}98.32
& \cellcolor{secondbest}0.366
& \cellcolor{best}0.001
& \cellcolor{secondbest}0.203 \\

\textbf{ROSIA}
& 70.1
& 69.15
& 0.532
& 0.696
& 0.328 \\

\textbf{Tetra}
& \cellcolor{secondbest}79.1
& \cellcolor{secondbest}91.83
& 0.593
& 6.053
& \cellcolor{best}0.183 \\

\bottomrule
\end{tabular}
\vspace{-4mm}
\end{table}

\subsection{Real-World Dataset}

\begin{figure*}[t]
  \centering
  \includegraphics[width=\linewidth]{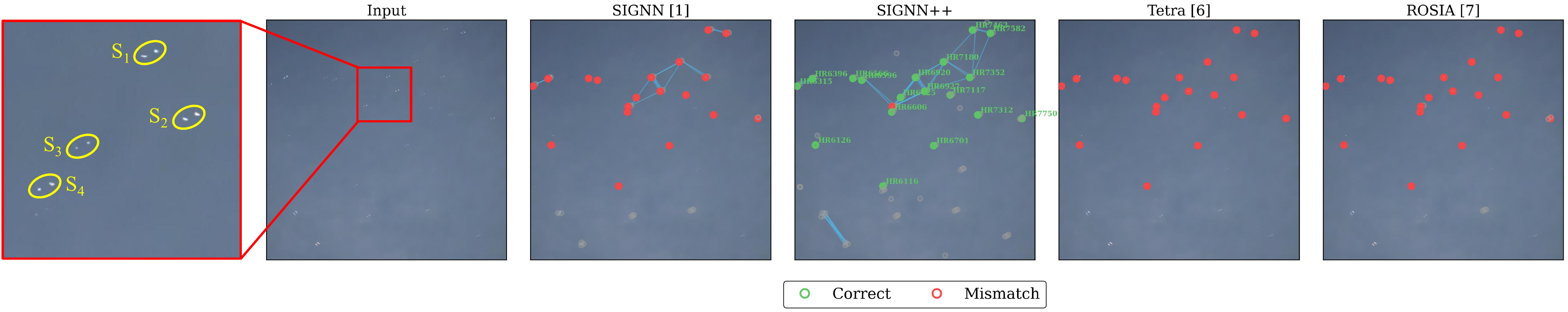}
  \vspace{-8mm}
  \caption{\textbf{Inference results on random-view real night-sky images.} The image contains overall motion blur caused by camera shake. SIGNN++ successfully solves the image. Since SIGNN++ connects edges only between stars separated by $1^\circ$--$4^\circ$, it is robust to duplicated centroids caused by motion blur and can successfully identify the stars.}
  \label{fig:random_image_inference}
\vspace{-4mm}
\end{figure*}

\begin{figure}[t]
  \centering
  \includegraphics[width=\linewidth]{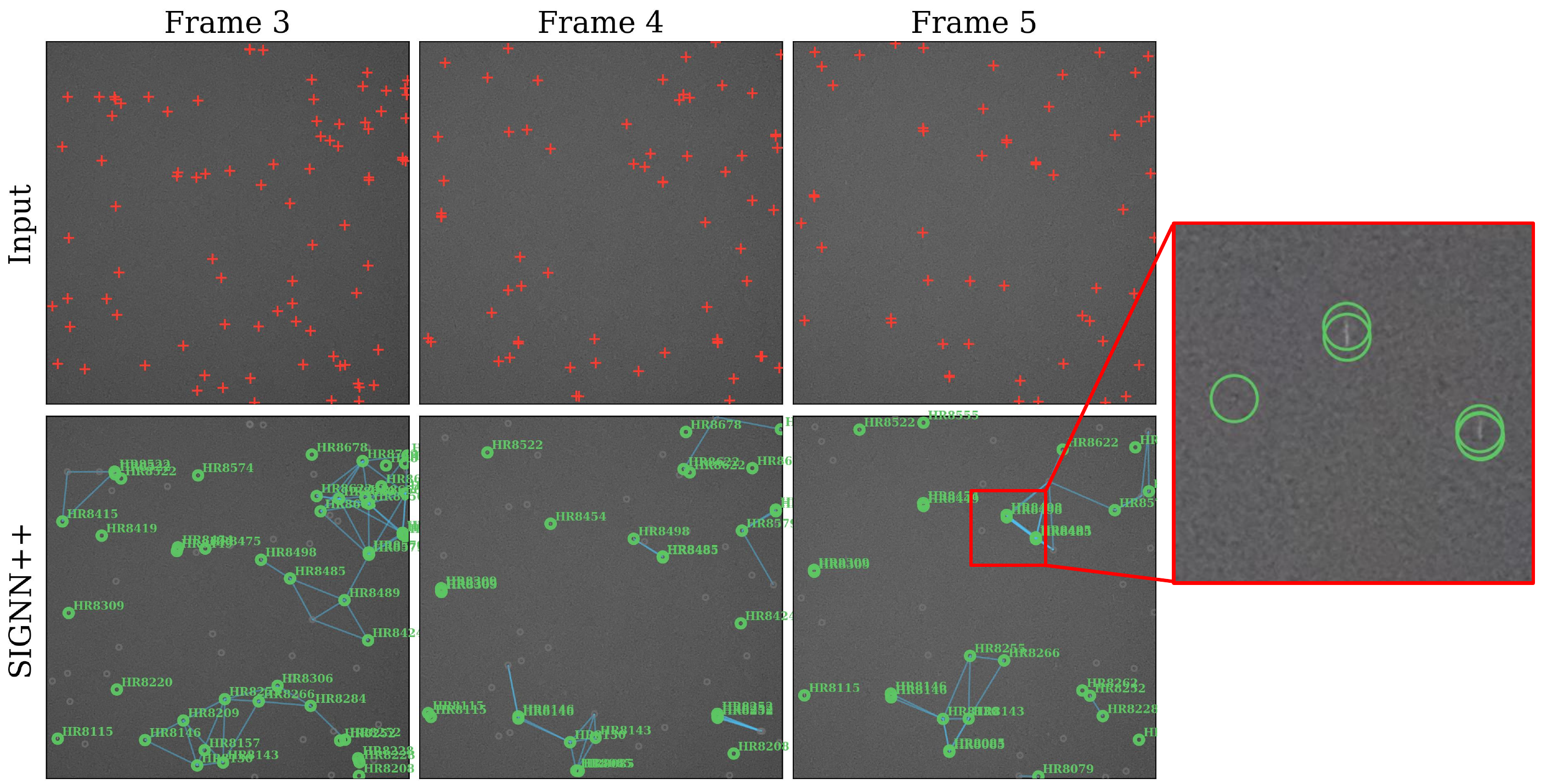}
  \vspace{-6mm}
  \caption{\textbf{Inference results on sequential real night-sky images.} SIGNN++ successfully solves six frames of the fifth sequence in Table~\ref{tab:sequence_performance}. As shown in the cropped image on the right, a star streak is detected as two separate centroids, causing the other SID methods to fail identification. In contrast, SIGNN++ successfully identifies the stars despite this duplicated centroid caused by the streak.}
  \label{fig:sequence_inference}
\vspace{-6mm}
\end{figure}

Table~\ref{tab:real_random_view} shows the results on the random-view dataset. The overall trend resembles the Stellarium results. SIGNN++ achieves the best solved rate and F1 score, SIGNN shows the shortest computation time, and Tetra achieves the lowest attitude error. Although Tetra appears to outperform SIGNN++ in terms of attitude error, the difference is not statistically significant because both methods show a standard deviation of approximately \(0.15^\circ\). Moreover, Tetra calculates the attitude error only from solved frames, whereas SIGNN++ includes more challenging frames where it still attempts attitude estimation. Therefore, this comparison is not fully favorable to SIGNN++. To summarize, SIGNN++ demonstrates comparable or better performance than the comparison methods across all evaluation metrics.

The benefit of the joint refinement is verified by the reprojection error.
PSA reduces the reprojection error from \(8.11 \pm 3.68\)~px to \(4.77 \pm 3.96\)~px on the random-view dataset by jointly optimizing the FOV and \(R\).
The remaining reprojection error is likely caused by unmodeled imaging effects, including camera distortion and centroiding errors, since we do not perform additional camera calibration.

Compared with SIGNN, SIGNN++ increases computation time due to PSA but substantially improves identification performance. Although the PSA module increases the computation time of SIGNN++ by approximately 0.35~s on average, it substantially improves the solved rate. Furthermore, SIGNN++ still requires less computation time than Tetra and ROSIA, demonstrating that the extension from SIGNN to SIGNN++ provides a worthwhile improvement.

Fig.~\ref{fig:random_image_inference} shows the identification results of SIGNN, SIGNN++, Tetra, and ROSIA for a random-view image. This image contains severe motion blur, causing the conventional SID methods to interpret duplicated centroids as false stars and preventing correct identification when the false detections exceed 50\%. In contrast, SIGNN++ connects only edges between stars separated by \(1^\circ\)--\(4^\circ\), preventing edge connections between duplicated centroids and enabling correct identification. In addition, SIGNN++ remains robust in images with cloud occlusion, where numerous false stars are generated. Therefore, SIGNN++ demonstrates robustness to severe blur and cloud-induced false detections in real terrestrial night-sky images.

Fig.~\ref{fig:sequence_inference} shows the SIGNN++ identification results for the fifth sequential image set. The top row shows the original images, while the bottom row visualizes the identification results as graphs overlaid on the images. Since the camera rotates in the pitch direction at an angular velocity of \(0.6^\circ/s\), vertical star streaks appear in the images. As shown in the cropped image on the right, the streaks are detected as two separate centroids, which directly degrades the performance of other SID methods. However, SIGNN++ remains robust to these duplicated centroids due to the edge connection strategy described above.

\begin{table}[h]
\vspace{-4mm}
\centering
\caption{Performance comparison on sequential real images.}
\vspace{-2mm}
\label{tab:sequence_performance}
\scriptsize
\setlength{\tabcolsep}{2pt}
\renewcommand{\arraystretch}{0.90}

\resizebox{\columnwidth}{!}{%
\begin{tabular}{c l l c c c c}
\toprule
\makecell[c]{\textbf{Seq.}} &
\makecell[c]{\textbf{Setup}} &
\makecell[l]{\textbf{Method}} &
\makecell[c]{\textbf{Solved}\\\textbf{Rate (\%)}} &
\makecell[c]{\textbf{Att. Error}\\\textbf{(deg)}} &
\makecell[c]{\textbf{Time Solved}\\\textbf{(s)}} &
\makecell[c]{\textbf{Time Failed}\\\textbf{(s)}} \\
\midrule

\multirow{3}{*}{\textbf{1}}
& \multirow{3}{*}{$90^\circ$, roll, $1^\circ$/s}
& \textbf{SIGNN++} & \cellcolor{best}100.000 & \cellcolor{best}0.137 & \cellcolor{best}0.071 & -- \\
& & \textbf{Tetra} & \cellcolor{best}100.000 & \cellcolor{secondbest}0.151 & \cellcolor{secondbest}0.428 & -- \\
& & \textbf{ROSIA} & \cellcolor{secondbest}88.889 & 0.222 & 0.568 & 1.698 \\
\midrule

\multirow{3}{*}{\textbf{2}}
& \multirow{3}{*}{$90^\circ$, roll, $2^\circ$/s}
& \textbf{SIGNN++} & \cellcolor{best}66.667 & \cellcolor{best}0.206 & \cellcolor{best}0.040 & 0.014 \\
& & \textbf{Tetra} & \cellcolor{secondbest}33.333 & \cellcolor{secondbest}0.404 & 0.755 & 4.477 \\
& & \textbf{ROSIA} & \cellcolor{secondbest}33.333 & 0.483 & \cellcolor{secondbest}0.356 & 0.636 \\
\midrule

\multirow{3}{*}{\textbf{3}}
& \multirow{3}{*}{$90^\circ$, roll, $3^\circ$/s}
& \textbf{SIGNN++} & \cellcolor{best}14.286 & -- & \cellcolor{best}0.024 & 0.013 \\
& & \textbf{Tetra} & \cellcolor{secondbest}0.000 & -- & -- & 4.391 \\
& & \textbf{ROSIA} & \cellcolor{best}14.286 & -- & \cellcolor{secondbest}0.817 & 0.247 \\
\midrule

\multirow{3}{*}{\textbf{4}}
& \multirow{3}{*}{$90^\circ$, pitch, $0.3^\circ$/s}
& \textbf{SIGNN++} & \cellcolor{best}100.000 & 0.194 & \cellcolor{best}0.198 & -- \\
& & \textbf{Tetra} & \cellcolor{secondbest}66.667 & \cellcolor{best}0.054 & \cellcolor{secondbest}0.239 & \textcolor{red}{5.385} \\
& & \textbf{ROSIA} & 33.333 & \cellcolor{secondbest}0.125 & 0.756 & 0.705 \\
\midrule

\multirow{3}{*}{\textbf{5}}
& \multirow{3}{*}{$90^\circ$, pitch, $0.6^\circ$/s}
& \textbf{SIGNN++} & \cellcolor{best}66.667 & \cellcolor{best}0.140 & \cellcolor{best}0.094 & 0.014 \\
& & \textbf{Tetra} & \cellcolor{secondbest}11.111 & -- & 0.231 & 4.863 \\
& & \textbf{ROSIA} & \cellcolor{secondbest}11.111 & -- & \cellcolor{secondbest}0.223 & 0.607 \\
\midrule

\multirow{3}{*}{\textbf{6}}
& \multirow{3}{*}{$90^\circ$, pitch, $1^\circ$/s}
& \textbf{SIGNN++} & \cellcolor{best}11.111 & -- & \cellcolor{best}0.014 & 0.007 \\
& & \textbf{Tetra} & \cellcolor{secondbest}0.000 & -- & -- & 1.864 \\
& & \textbf{ROSIA} & \cellcolor{best}11.111 & -- & \cellcolor{secondbest}0.034 & 0.051 \\
\midrule

\multirow{3}{*}{\textbf{7}}
& \multirow{3}{*}{$20^\circ$, roll, $1^\circ$/s}
& \textbf{SIGNN++} & \cellcolor{best}44.444 & \cellcolor{best}0.065 & \cellcolor{best}0.049 & 0.017 \\
& & \textbf{Tetra} & 0.000 & -- & -- & 4.735 \\
& & \textbf{ROSIA} & \cellcolor{secondbest}11.111 & -- & \cellcolor{secondbest}1.343 & 1.491 \\
\midrule

\multirow{3}{*}{\textbf{8}}
& \multirow{3}{*}{$20^\circ$, roll, $2^\circ$/s}
& \textbf{SIGNN++} & \cellcolor{best}22.222 & \cellcolor{best}0.091 & \cellcolor{best}0.023 & 0.011 \\
& & \textbf{Tetra} & 0.000 & -- & -- & 3.221 \\
& & \textbf{ROSIA} & \cellcolor{secondbest}11.111 & -- & \cellcolor{secondbest}0.844 & 0.588 \\
\midrule

\multirow{3}{*}{\textbf{9}}
& \multirow{3}{*}{$20^\circ$, roll, $3^\circ$/s}
& \textbf{SIGNN++} & \cellcolor{best}11.111 & -- & \cellcolor{best}0.021 & 0.017 \\
& & \textbf{Tetra} & \cellcolor{secondbest}0.000 & -- & -- & 3.622 \\
& & \textbf{ROSIA} & \cellcolor{best}11.111 & -- & \cellcolor{secondbest}0.624 & 1.110 \\
\midrule

\multirow{3}{*}{\textbf{10}}
& \multirow{3}{*}{$50^\circ$, roll, $1^\circ$/s}
& \textbf{SIGNN++} & \cellcolor{best}100.000 & \cellcolor{secondbest}0.139 & \cellcolor{secondbest}0.404 & -- \\
& & \textbf{Tetra} & \cellcolor{best}100.000 & \cellcolor{best}0.131 & \cellcolor{best}0.298 & -- \\
& & \textbf{ROSIA} & \cellcolor{secondbest}55.556 & 0.340 & 0.687 & 1.020 \\
\midrule

\multirow{3}{*}{\textbf{11}}
& \multirow{3}{*}{$70^\circ$, roll, $1^\circ$/s}
& \textbf{SIGNN++} & \cellcolor{best}66.667 & \cellcolor{best}0.180 & \cellcolor{best}0.031 & 0.016 \\
& & \textbf{Tetra} & \cellcolor{secondbest}11.111 & -- & 3.220 & \textcolor{red}{5.019} \\
& & \textbf{ROSIA} & \cellcolor{secondbest}11.111 & -- & \cellcolor{secondbest}1.246 & 0.958 \\
\midrule

\multirow{3}{*}{\textbf{12}}
& \multirow{3}{*}{$20^\circ$, roll, $1^\circ$/s}
& \textbf{SIGNN++} & \cellcolor{best}66.667 & \cellcolor{best}0.067 & \cellcolor{best}0.020 & 0.012 \\
& & \textbf{Tetra} & 11.111 & -- & \cellcolor{secondbest}0.165 & 3.987 \\
& & \textbf{ROSIA} & \cellcolor{secondbest}33.333 & \cellcolor{secondbest}0.488 & 0.179 & 0.076 \\
\midrule

\multirow{3}{*}{\textbf{13}}
& \multirow{3}{*}{$40^\circ$, roll, $1^\circ$/s}
& \textbf{SIGNN++} & \cellcolor{best}100.000 & \cellcolor{best}0.116 & \cellcolor{best}0.108 & -- \\
& & \textbf{Tetra} & \cellcolor{best}100.000 & \cellcolor{secondbest}0.120 & \cellcolor{secondbest}0.113 & -- \\
& & \textbf{ROSIA} & \cellcolor{secondbest}66.667 & 0.288 & 0.689 & 0.912 \\
\midrule

\multirow{3}{*}{\textbf{14}}
& \multirow{3}{*}{$30^\circ$, roll, $0.3^\circ$/s}
& \textbf{SIGNN++} & \cellcolor{best}88.889 & \cellcolor{best}0.089 & \cellcolor{best}0.039 & 0.017 \\
& & \textbf{Tetra} & \cellcolor{secondbest}77.778 & \cellcolor{secondbest}0.100 & \cellcolor{secondbest}0.508 & \textcolor{red}{6.634} \\
& & \textbf{ROSIA} & \cellcolor{secondbest}77.778 & 0.352 & 0.884 & 1.897 \\
\midrule

\multirow{3}{*}{\textbf{15}}
& \multirow{3}{*}{$50^\circ$, pitch, $0.3^\circ$/s}
& \textbf{SIGNN++} & \cellcolor{best}100.000 & \cellcolor{best}0.046 & \cellcolor{best}0.122 & -- \\
& & \textbf{Tetra} & \cellcolor{secondbest}88.889 & \cellcolor{secondbest}0.049 & 1.003 & 1.178 \\
& & \textbf{ROSIA} & 66.667 & 0.164 & \cellcolor{secondbest}0.537 & 0.614 \\

\bottomrule
\end{tabular}
}
\vspace{-2mm}
\end{table}

Table~\ref{tab:sequence_performance} presents the evaluation metrics of SIGNN++, ROSIA, and Tetra for each sequence.
The setup column lists elevation, rotation direction, and angular velocity in that order.

Overall, SIGNN++ shows better performance in solved rate, attitude error, and time solved. Among the 15 sequences, SIGNN++ achieves the highest solved rate in all sequences, ranks first in attitude error for 11 sequences, and ranks first in computation time for 14 sequences. In particular, for the 5th, 7th, and 11th sequences, SIGNN++ successfully solves some frames, whereas Tetra and ROSIA fail on all frames. A common characteristic of these sequences is the presence of partial cloud occlusion and star streaks. This observation is consistent with the robustness of SIGNN++ to clouds and star streaks demonstrated in the random-view image analysis.

Another advantage of SIGNN++ can be seen from the values shown in red: Tetra requires more than 5~s for images in which identification fails. Therefore, it is difficult to guarantee real-time positioning when using Tetra. Although the positioning is performed separately after image acquisition and is therefore not fully real-time, the computation time indicates that SIGNN++ generally achieves a high frame solved rate and performs positioning within 0.3~s, making it the most suitable for real-time operation among the evaluated SID methods.

Compared with the computation time of SIGNN++ on the random-view images, the sequential images are up to approximately 10 times faster. This is likely because the exposure time is reduced to nearly one-third, resulting in significantly fewer detected stars and a substantial reduction in the time required for PSA. We do not quantitatively analyze the correlation between the number of observed stars and computation time. However, since PSA requires more computation time than the network inference itself, the reduction in PSA processing time is considered the main reason for the overall decrease in computation time.

When grouping sequences with the same elevation, i.e., \{1, 2, 3\}, \{4, 5, 6\}, and \{7, 8, 9\}, the other SID methods almost completely fail to identify stars as the angular velocity increases, whereas SIGNN++ still maintains non-zero solved rates. In addition, comparing sequences \{1, 10, 7\}, where only the elevation changes from \(90^\circ\) to \(50^\circ\) and \(20^\circ\), shows that the solved rates of the other algorithms decrease significantly as the elevation decreases, while SIGNN++ still identifies stars in approximately half of the frames. These results further indicate the robustness of SIGNN++ to atmospheric refraction and star streaks. However, the number of sequences remains limited, and more experiments are required to fully validate its generalization capability.

\subsection{Limitation}

First, SIGNN++ requires an approximate initial FOV, although this is considered an acceptable assumption because the FOV is determined by the camera intrinsics. Second, the irregular ST delay and its physical measurement period of more than 1~s limit the real-world experiments, which prevents exact synchronization with image acquisition. Third, the camera motion is limited to constant-velocity motion, making it difficult to generalize the current results to fully real-time scenarios.

%% file: Conclusion.tex
\vspace{-2mm}
\section{Conclusion}
\label{sec:Conclusion}

We propose SIGNN++, which extends SIGNN with three major improvements: a more realistic terrestrial noise model, Fourier-transformed edge features, and a RANSAC-based Perspective Star Alignment for attitude refinement. Through extensive evaluations on simulated and real terrestrial datasets, SIGNN++ demonstrates improved robustness to observation noise, light pollution, and challenging ground-based conditions. Compared with the original SIGNN and publicly available representative SID methods, SIGNN++ achieves more robust performance in real terrestrial environments. Future work will focus on further validating SIGNN++ through real-time experiments on mobile platforms.

\vspace{-4mm}